\documentclass[10pt, conference, compsocconf]{IEEEtran}
\usepackage{url}
\usepackage{etoolbox}
\makeatletter
\patchcmd{\@makecaption}
  {\scshape}
  {}
  {}
  {}
\makeatletter
\patchcmd{\@makecaption}
  {\\}
  {.\ }
  {}
  {}
\makeatother
\def\tablename{Table}

\usepackage{qcircuit}
\usepackage{svg}
\svgsetup{inkscapeopt=-C}
\usepackage{tikz}
\usepackage{array}
\usepackage{multirow}
\usepackage{pgfplots}

\usetikzlibrary{patterns}
\usetikzlibrary{arrows}
\usetikzlibrary{external}
\pgfplotsset{{compat=1.14}}

\usepackage{soul}
\usepackage[most]{tcolorbox}

\begin{document}
\title{Let Each Quantum Bit Choose Its Basis Gates}

\author{
\IEEEauthorblockN{Sophia Fuhui Lin\IEEEauthorrefmark{1},
Sara Sussman\IEEEauthorrefmark{2},
Casey Duckering\IEEEauthorrefmark{1}, 
Pranav S. Mundada\IEEEauthorrefmark{3},\\
Jonathan M. Baker\IEEEauthorrefmark{1},
Rohan S. Kumar\IEEEauthorrefmark{1},
Andrew A. Houck\IEEEauthorrefmark{3},
Frederic T. Chong\IEEEauthorrefmark{1}
}

\IEEEauthorblockA{\IEEEauthorrefmark{1}Department of Computer Science, University of Chicago, Chicago, USA\\
Email: \{sophialin1, cduck, jmbaker, rohansk\}@uchicago.edu, chong@cs.uchicago.edu} 
\IEEEauthorblockA{\IEEEauthorrefmark{2}Department of Physics, Princeton University, Princeton, USA\\
Email: sarafs@princeton.edu}
\IEEEauthorblockA{\IEEEauthorrefmark{3}Department of Electrical Engineering, Princeton University, Princeton, USA\\
Email: pmundada@alumni.princeton.edu, aahouck@princeton.edu}

}

\maketitle
\newtheorem{thm}{Theorem}[section]


\begin{abstract}
Near-term quantum computers are primarily limited by errors 
in quantum operations (or {\it gates}) between two quantum bits (or {\it qubits}).
A physical machine typically provides
a set of basis gates that include primitive 2-qubit (2Q) and 1-qubit (1Q)
gates that can be implemented in a given technology. 2Q {\it entangling} gates, coupled with some 1Q gates, allow for universal quantum computation. In superconducting
technologies, the current state of the art is to implement the same 2Q
gate between every pair of qubits (typically an XX- or XY-type gate).  This strict hardware uniformity requirement for 2Q gates in a large quantum computer has made scaling up a time and resource-intensive endeavor in the lab.

We propose a radical idea -- allow the 2Q basis gate(s) to differ between every pair of qubits, selecting the best entangling gates that can be calibrated between given pairs of qubits. This work aims to give quantum scientists the ability to run meaningful algorithms with qubit systems that are not perfectly uniform. Scientists will also be able to use a much broader variety of novel 2Q gates for quantum computing.
We develop a theoretical framework for identifying good 2Q basis gates on ``nonstandard'' Cartan trajectories that deviate from ``standard'' trajectories like XX. We then introduce practical methods for calibration and compilation with nonstandard 2Q gates, and discuss possible ways to improve the compilation.
To demonstrate our methods in a case study, we simulated both standard XY-type trajectories and faster, nonstandard trajectories using an entangling gate architecture with far-detuned transmon qubits. We identify efficient 2Q basis gates on these nonstandard trajectories and use them to compile a number of standard benchmark circuits such as QFT and QAOA. Our results demonstrate an 8x improvement over the baseline 2Q gates with respect to speed and coherence-limited gate fidelity.
\end{abstract}
\IEEEpeerreviewmaketitle
\section{Introduction}\label{sec_intro}

Quantum computers have the potential to solve problems currently intractable for conventional computers \cite{shor1999polynomial}, but current computations are limited by errors \cite{preskill2018nisq}, particularly when interacting two qubits to perform a quantum gate operation. This is not surprising, as qubits are engineered to preserve quantum state and be isolated from the environment, but a quantum operation is a moment in time where external control is applied from the environment to deliberately alter a qubit's state.  To accomplish low-error gates, the control mechanisms are carefully designed and the control signals are calibrated for each qubit or pair of qubits.

\begin{figure}
    \centering
    \scalebox{0.6}{
    \includesvg{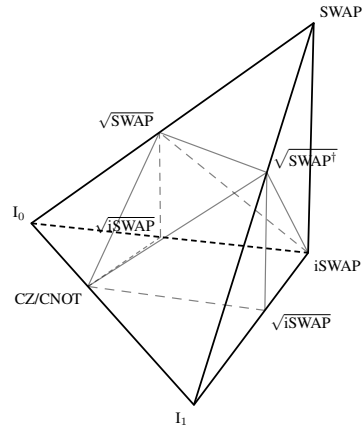}
    }
    \caption{The Weyl chamber of 2Q quantum gates, explained in Sec \ref{bg_kak}. The non-local part of a 2Q gate is fully described by its position in the Weyl chamber. As the duration of an entangling gate pulse increases, the 2Q gate evolves, traversing a Cartan trajectory in the Weyl chamber. CNOT and CZ are both represented by $(\frac{1}{2},0,0)$. The SWAP gate is at the top vertex $(\frac{1}{2},\frac{1}{2},\frac{1}{2})$. On the bottom surface, $(t_x,t_y,0)$ and $(1-t_x, t_y, 0)$ represent the same equivalent class of gates. For example, the two points $I_0 = (0,0,0)$ and $I_1 = (1,0,0)$ both represent the 2Q identity gate $I$.}
    \label{fig:weyl_chamber}
\end{figure}

Similar to how classical computers use a small set of classical logic gates (AND, OR, NOT, XOR...) as building blocks for larger circuits, current superconducting quantum devices typically only directly support a universal gate set consisting of a few two-qubit (2Q) gates and a continuous set of single-qubit (1Q) gates. This paper will refer to the set of directly supported quantum gates as \textit{basis gates}. In the space of 2Q gates (see Fig.\ref{fig:weyl_chamber}), any point that does not coincide with SWAP or Identity has nonzero entangling power. Any of these 2Q entangling gates can achieve universal computation when added to a continuous set of 1Q gates \cite{bremner2002practical}.

Using the minimal set of gates needed for universal computing is rarely a desirable thing to do. For example, while the NAND is universal in classical computing, building circuits from it alone is less efficient than using a larger set of logic gates. However, the intensive calibrations necessary for 
high fidelity 2Q gates between qubits in a large quantum computer make it impractical to support a large set of 2Q basis gates. All logical 2Q gates scheduled to run on a quantum computer have to be decomposed by its compiler into alternating layers of pre-calibrated 1Q and 2Q basis gates. Thus, the choice of which 2Q gates to directly support is critical to enabling high-performance quantum computing. On the hardware side, the 2Q basis gates must have high-fidelity hardware implementations. On the software side, they must enable the low-depth decomposition of other 2Q gates.

Superconducting qubits support XX- and XY-type 2Q interactions \cite{Abrams2020, Kwon2021}. The strength of each of these interactions depends on the type of coupling, the coupling strength, and the frequency detuning between the qubits \cite{Kwon2021}. The Weyl chamber space of 2Q gates (Fig. \ref{fig:weyl_chamber}) is a useful way to visualize these interactions: the coordinates of a gate correspond to its non-local part in Cartan's KAK decomposition (see Section \ref{bg_kak}). In the Weyl chamber, gates in the XX family form a straight trajectory from Identity to CNOT/CZ, while gates in the XY family form a trajectory from Identity to iSWAP. Cartan trajectories are generated by increasing the duration of an entangling gate pulse, which evolves the 2Q gate. 

Section \ref{deviation} describes the various difficulties associated with reliably performing standard 2Q gates like CZ and iSWAP on today's superconducting quantum computers. Whether a deviation from a standard Cartan trajectory is a 2Q error depends on what the target 2Q gate is. If the target 2Q gate has to be a certain standard gate (e.g. iSWAP), even a small amount of coherent crosstalk between the two qubits will cause the gate to have a small CZ component, and so the gate will not be identically iSWAP. However, if that coherent crosstalk is a stable systematic that does not add noise or cause decoherence, the gate could still be an effective, high fidelity entangler that is useful for computing; the target 2Q gate would just have to be a nonstandard unitary rather than the standard iSWAP. In Section \ref{deviation} we also show an example of an experimentally measured Cartan trajectory of 2Q gates on a superconducting qubit device. The measured Cartan trajectory includes very fast nonstandard 2Q gates with high entangling power, but it does not pass through traditional basis gates like iSWAP and CZ. This example motivates us to develop methods that enable the use of nonstandard 2Q gates for quantum computing. If we allow for deviations from standard 2Q gates, we can use high fidelity, non-standard quantum hardware for practical quantum computing.

Using nonstandard 2Q basis gates requires methods for identifying good basis gates on a general 2Q gate trajectory, calibrating a nonstandard gate, and compiling with nonstandard basis gates. The primary focus of our work is to construct and demonstrate a framework for efficiently identifying a ``good'' set of 2Q basis gates from a nonstandard trajectory. But we also propose solutions for calibration and compilation. 

What are our standards for a good set of 2Q basis gates? Following the principle of Amdahl’s Law, we pay most attention to the SWAP gate as a target for synthesis because of its importance for communication within programs executing on superconducting devices. To mitigate crosstalk and satisfy other hardware constraints, superconducting devices usually have the sparse connectivity of a grid lattice or a hexagonal lattice. Therefore, the compiler has to schedule a series of SWAP gates before it can interact two qubits that are not adjacent to each other. Although clever mapping from logical to physical qubits can result in a smaller number of inserted SWAP gates, we still observe a high proportion of SWAP gates in post-mapping quantum circuits. Besides efficient synthesis of the SWAP gate, our framework also allows one to prioritize other target gates, including but not limited to CNOT, iSWAP, and the B gate. It also enables the simultaneous prioritisation of multiple target gates.

The calibration of a non-standard 2Q basis gate requires identifying a gate duration that gives an ideal basis gate and then accurately characterizing the corresponding gate so that we use the right unitary for compiling. Our proposed calibration protocol address both without causing a long downtime on a quantum device. However, we point out that in order to \textit{precisely} characterize a non-standard gate, one should consider using gate set tomography (GST) as opposed to quantum process tomography (QPT). The data collected from GST experiments may require several hours of classical processing. Before that finishes, one would have to use the calibration results from the previous cycle. The speedup of GST's classical processing, which is an active field of research \cite{pygsti}, would help reduce the cost of calibrating non-standard gates. In addition, we observed that the systematic deviations are stable over days (Fig. \ref{fig:Expt_Fig2}). If the change in deviation is negligible, one may not need to apply GST in every calibration cycle.

Compiling with non-standard 2Q basis gates requires a conversion from arbitrary 2Q gates into the basis gates. There isn't a general analytical formula that works for arbitrary target and basis gate, so a numerical search is needed. However, we can analytically obtain information on the minimum circuit depth needed for a perfect synthesis and use it to facilitate the numerical search. Besides, the circuits that synthesise common gates from the basis gates can be pre-computed after each calibration cycle, so that one wouldn't need to re-compute them for every program.

Our contributions are summarized below.
\begin{itemize}
    \item 
    Our work is the first to consider using 2Q basis gates from general non-standard gate trajectories that are not parametrized by a simple function.
    \item We provide a theoretical framework for identifying and visualizing the set of good 2Q basis gates, given a set of target 2Q gates to prioritize. With an emphasis on SWAP, we characterize the sets of gates that enable the synthesis of SWAP in 1, 2, and 3 layers, respectively. As another example, we visualize the gates that are able to both synthesize SWAP in 3 layers and CNOT in 2 layers. After identifying the volume of desirable basis gates in the Weyl chamber, one can select the first intersection of the trajectory with the volume as the 2Q basis gate. (Section \ref{sec_theory})
    \item We propose a practical calibration protocol that is agnostic as to whether a 2Q gate is standard or non-standard. (Section \ref{sec_calibration})
    \item We discuss a practical approach to compiling with non-standard 2Q basis gates. (Section \ref{sec_compile})
    \item We apply our methods to a case study entangling gate architecture with far-detuned transmon qubits \cite{hamiltonian_source}. First, we use our theoretical framework to select 2Q basis gates from simulated nonstandard Cartan trajectories that are realistic for this case study architecture. By increasing the entangling pulse drive amplitude we get a significant 2Q basis gate speedup but introduce a deviation into the Cartan trajectory. Then we use these 2Q basis gates to run a variety of benchmark circuits including BV\cite{BV}, QAOA\cite{qaoa}, the QFT adder\cite{qft}, and the Cuccaro Adder\cite{cuccaro2004new}, and compare to the results from using the $\sqrt{iSWAP}$ gate on the standard XY-type trajectory. (Section \ref{case_study})
\end{itemize}

\section{Background}\label{sec_bg}
\subsection{Qubits and gates}
Unlike a classical bit that is either 0 or 1, a quantum bit (qubit) can exist in a linear superposition of $|0\rangle$ and $|1\rangle$; A general quantum state can be expressed as $\alpha |0\rangle + \beta |1\rangle$ where $\alpha, \beta$ are complex amplitudes that satisfy $|\alpha|^2 + |\beta|^2 = 1$. Thus, the state of one qubit can be represented by a 2-vector of the amplitudes $\alpha$ and $\beta$. A system of $n$ qubits can exist in a superposition of up to $2^n$ basis states, and its state can be represented by a $2^n$-vector of complex amplitudes. A quantum gate that acts on $n$ qubits can be represented by a $2^n \times 2^n$ unitary matrix. 

\subsection{Geometric characterization of 2Q gates}\label{bg_kak}
Two 2Q quantum gates $U_1, U_2 \in SU(4)$ are \textit{locally equivalent} if it is possible to obtain one from the other by adding 1Q operations. In other words, 2Q operations $U_1$ and $U_2$ are locally equivalent if there exist $k_1, k_2 \in SU(2) \otimes SU(2)$ such that $U_1 = k_1 U_2 k_2$. For example, CNOT and CZ are locally equivalent via Hadamard gates.

Any 2Q quantum gate $U\in SU(4)$ can be written in the form of
\begin{equation}\label{kak}
    U = k_1 \exp (-i\frac{\pi}{2} (t_x X\otimes X + t_y Y\otimes Y + t_z Z\otimes Z )) k_2
\end{equation}
 where $X, Y, Z$ are the Pauli gates. This is called the Cartan decomposition.
 
The space of two-qubit quantum gates can be represented geometrically in a Weyl chamber (Fig. \ref{fig:weyl_chamber}), where each point stands for a set of gates that are locally equivalent to each other \cite{zhang2003geometric}. The Cartan coordinates $(t_x, t_y, t_z)$ in Eq. \eqref{kak} are the coordinates of $U$ in the Weyl chamber. They fully characterize the \textit{non-local} part of a 2Q gate. On the bottom surface, $(t_x,t_y,0)$ and $(1-t_x, t_y, 0)$ represent the same equivalent class of gates. The other points in the Weyl chamber each represent a different equivalence class of 2Q gates. We refer the interested readers to \cite{crooks_tutorial} for a more thorough introduction to the Weyl chamber. Note that other conventions of the Cartan coordinates are also common. They usually differ from ours by a constant factor of $\pi$ or $2\pi$.

In this paper, when we talk about some gate $G$ in the Weyl chamber, we usually mean the local equivalence class of 2Q gates that includes $G$.

\subsection{Entangling power of 2Q gates}\label{background_ep}
The entangling power \cite{entanglingpowerdef} is a widely accepted quantitative measure of the capacity of a 2Q gate to entangle the qubits that it acts on. It is typically a good indicator of the ability of a specific 2Q gate to synthesize arbitrary 2Q gates. For a unitary operator $U$, the entangling power $e_p(U) \in [0, \frac{2}{9}]$ is defined as the average linear entropy of the states produced by $U$ acting on the manifold of all separable states \cite{entanglingpowerdef}. It is solely based on the non-local part of $U$, which is characterized by the position of $U$ in the Weyl chamber.

A 2Q gate has 0 entangling power if and only if it is locally equivalent to the Identity or the SWAP gate. Conversely, 2Q gate $U$ is called a \textit{perfect entangler} if it can produce a maximally entangled state from an unentangled one\cite{zhang2003geometric}. Perfect entanglers (PE) have entangling power no less than $\frac{1}{6}$. They constitute a polyhedron in the Weyl chamber that is exactly half of the total volume. The 6 vertices of the PE polyhedron are CZ(CNOT), iSWAP, $\sqrt{SWAP}$, $\sqrt{SWAP}^\dagger$, and the 2 points that both represent $\sqrt{iSWAP}$. The perfect entanglers with maximal entangling power of $\frac{2}{9}$ are also called \textit{special perfect entanglers}\cite{pe2004}. In the Weyl chamber, they are on the line segment from CNOT to iSWAP. The B gate, which is at the midpoint of this line segment, has the property that it can synthesize any arbitrary 2Q gates within 2 layers\cite{bgate}. However, there has been no proposal to directly implement the B gate in hardware.


\section{Related work}\label{sec_related}

To the best of our knowledge, no prior work involves using 2Q basis gates from arbitrary nonstandard gate trajectories. In parallel with this work, Lao et al. \cite{lao2022software} propose to mitigate coherent parasitic errors in 2Q gates by software and present methods of compilation. Our work is more general then \cite{lao2022software}, although we share the insight that coherent errors in 2Q gates can be treated as part of the gate for compilation. While our framework works for general irregular trajectories and select basis gates on them using the approach detailed in Section \ref{sec_theory}, they focus on iSWAP-like (XY) gates with an unwanted CPHASE (XX) component (which belongs to the FSim gate set so is not truly non-standard) and always use CPHASE($\psi$)iSWAP($\pi$/4) because it has similar expressivity as iSWAP($\pi$/4) for small deviation $\psi$. They do not discuss calibration. Their baseline for evaluation is similar to the baseline in our case study, which is to make the trajectory more standard by lengthening the gate duration.

Recent research from both the experimental \cite{foxen,moskalenko_cont2Q,xiong_cont2Q,reagor_cont2Q} and theory sides has utilized 2Q (and 3Q) basis gates from a continuous set of standard gates, as opposed to only building and compiling with the best-known gates like CNOT and iSWAP. The works that are most relevant to this project are those that look for a small set of 2Q basis gates (from a continuous standard gate set) that are the most valuable to calibrate. Lao et al. \cite{prakash} use a numerical approach to test the performance of different gates from the fSim and XY gate sets on a range of application circuits, with the overall circuit success rate as the objective. Peterson el al. \cite{peterson2021xx} from IBM use analytic techniques to find that the gate set $\{CX, CX^{1/2}, CX^{1/3}\}$ is almost as good as the entire continuous set of XX gates in implementing random 2Q gates. They try to minimize the expected (average) infidelity in implementing random 2Q gates under an experimentally motivated error model. Huang et al. \cite{sqiswap} proposes using the $\sqrt{iSWAP}$ as 2Q basis gate, instead of using iSWAP or CNOT, and implement it using a 2-fluxonium qubit device. Recent proposals for novel nonstandard 2Q gates in the superconducting qubit literature that are informed by the current experimental challenges in scaling up with standard 2Q gates include \cite{perez_nonstandard2Qgate,xu_nonstandard2Qgate}.

\section{Systematic deviations in 2Q \\ gates}\label{deviation}

The 2Q gate is a critical building block that must be well-engineered before it is used to construct a quantum computer with many qubits. In practice, engineering 2Q gates in the lab involves iterating prototypes of the devices to minimize any and all systematic errors that result from imperfect device design or control along with nonuniformities in device fabrication.

Even if unwanted crosstalk between the qubits is successfully reduced and the 2Q gate is shown to be an effective entangler with a consistent identity, if the gate's identity is somehow nonstandard, one would normally assume it is not useful. The constraint of requiring 2Q gates to be standard is most burdensome for the superconducting qubit platform, where device Hamiltonians are engineered from scratch and there is no 2Q gate that is truly native to the platform - unlike, for instance, the SWAP gates that are native to atomic qubits \cite{Vool2017}.

Today's multi-qubit superconducting devices are not able to perform perfectly identical 2Q gates between every pair of qubits because of device-level imperfections, tradeoffs and uncertainties. Experimentalists model the expected rate of information leakage between on-chip elements using microwave circuit design software \cite{microwaveoffice, ansys}, but it is inevitable that irregularities arise during device fabrication and packaging. The devices are at least partially handmade and every fabrication tool has a finite precision. Also, the various materials that make up the layers of the superconducting device can host physical two-level systems that act as sources of noise and even can coherently interact with qubits \cite{tls1, tls2}; reducing the effect of these two-level systems is an active field of research \cite{tls3}. Another active field of research is reducing irregularities in the fabrication of Josephson junctions, which are critical on-chip elements \cite{lbnl, qubit_freq_std}. For a given device, it can be difficult for the experimentalist to determine whether a systematic 2Q gate deviation is caused by an imperfection in the device design or in its control. For example, a common source of systematic 2Q gate deviation is the imperfect mitigation of the static ZZ crosstalk which is a dominant source of 2Q gate error for transmon qubits \cite{mundada_et_al_2019,ku2020suppression,sung2020realization,noguchi2020fast,kandala2020demonstration,zhao2020suppression}. Devices can be designed to suppress the static ZZ crosstalk but unless the device is properly fabricated, packaged, biased and controlled there will be nonzero static ZZ crosstalk which will cause the 2Q gate to deviate from the target unitary.

Superconducting devices can also have higher order Hamiltonian terms that result in the experimentally measured Cartan trajectory of 2Q gates deviating from the expected Cartan trajectory. This deviation is particularly significant for fast gates enabled by large coupling or large drive strength \cite{hamiltonian_source, McKay2016, jurcevic2021demonstration}. Experimentalists have historically tried to suppress these deviations by reducing the 2Q gate drive strength, which has the negative consequence of slowing the 2Q gate down. It is in general difficult to accurately model the effect of the strong drives that perform fast 2Q gates on the Hamiltonian level, and this is an active field of research \cite{hamiltonian_source}.

\begin{figure}[!ht]
    \centering
    \includegraphics[width=0.99\columnwidth]{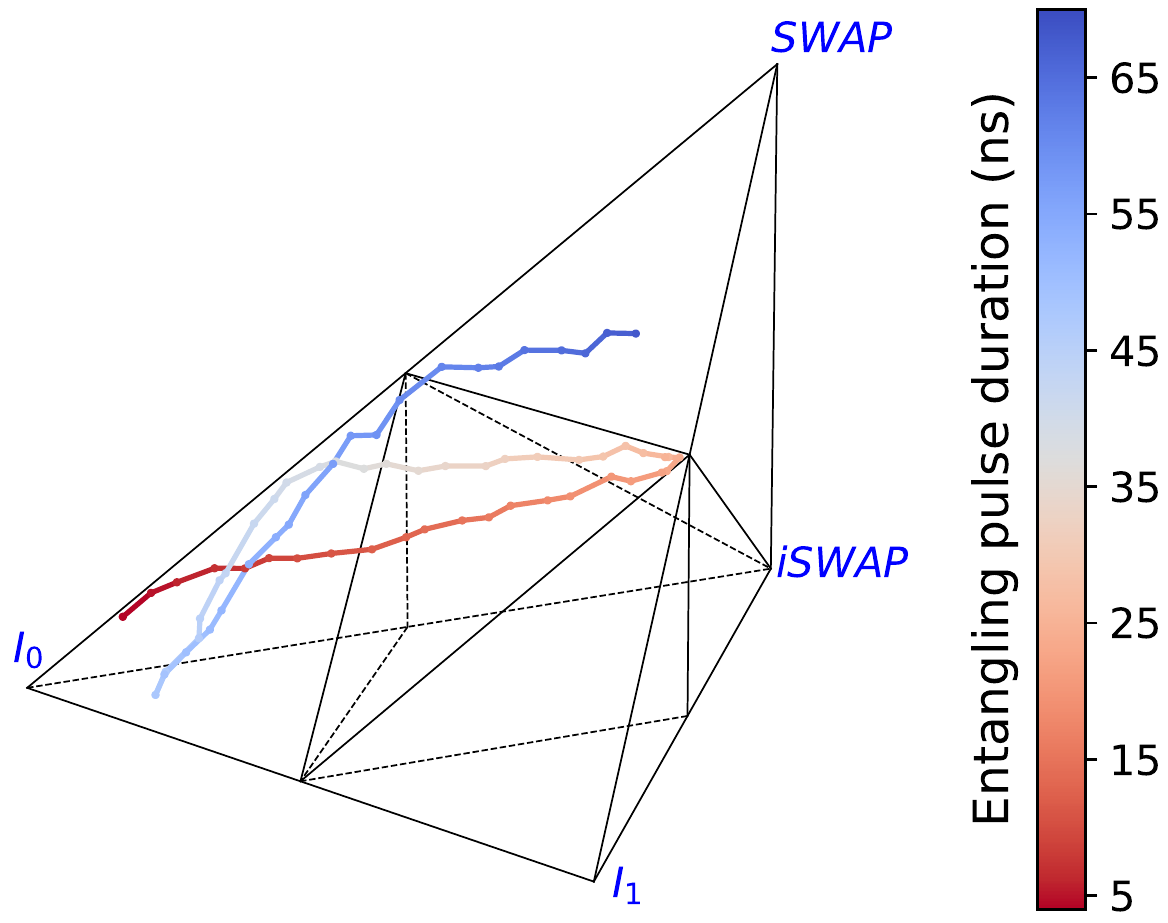}
    \caption{Experimental data showing a nonstandard Cartan coordinate trajectory. An experimental implementation \cite{experimental_work} of the iSWAP gate with the entangler architecture proposed in \cite{hamiltonian_source} yielded a nonstandard Cartan coordinate trajectory close to the plane of $I_0$, SWAP, and iSWAP. The first instance of a perfect entangler was at an entangler duration of 13 ns. In this nonstandard trajectory, the 13 ns entangler is offset from the Cartan coordinate for the square root of iSWAP and the 26 ns entangler is likewise offset from the Cartan coordinate for iSWAP. Note that due to an experimental hardware constraint the shortest possible entangling pulse duration was 4 ns, so the measured Cartan trajectory begins there.}
    \label{fig:Expt_Fig1}
\end{figure}

Plotting measured 2Q gates in Cartan coordinates is a valuable tool experimentalists can use to easily visualize and study any deviations their gates may have from the expected Cartan trajectory. For example, Figure \ref{fig:Expt_Fig1} shows a measured Cartan trajectory that is nonstandard. This experimental data was collected from one of the first iterations of a superconducting device \cite{experimental_work} that was designed to implement a recently proposed entangling gate architecture \cite{hamiltonian_source}. The data includes a very fast (13 ns) perfect entangler. 
Since the measured trajectory was systematically offset from the predicted one (XY), the experimentalists investigated potential sources of that systematic offset. Since this source of deviation could be eliminated with better device and control engineering, the experimentalists began to optimize their next device iteration accordingly. But in this work we suggest that there is nothing inherently unusable about measured Cartan trajectories that are nonstandard due to this kind of coherent systematic offset, and that the 13 ns nonstandard perfect entangler identified in Figure \ref{fig:Expt_Fig1} could be treated as a native 2Q basis gate by the compiler.

Our work seeks to enable the use of the nonstandard 2Q gates that can be native to superconducting devices. 

If 2Q gate calibration and compiling protocols became more flexible, usable superconducting 2Q gate yield would increase considerably, enabling more rapid and effective prototyping of 2Q gates which could be scaled to a computer. Furthermore, any number of novel superconducting devices with very fast 2Q gates that happen to be nonstandard could be effectively utilized for computing.

\section{Identifying good 2Q basis gates}\label{sec_theory}

\subsection{Fidelity of a synthesized gate}

\begin{figure*}
    \centering
    \includegraphics[width=\textwidth]{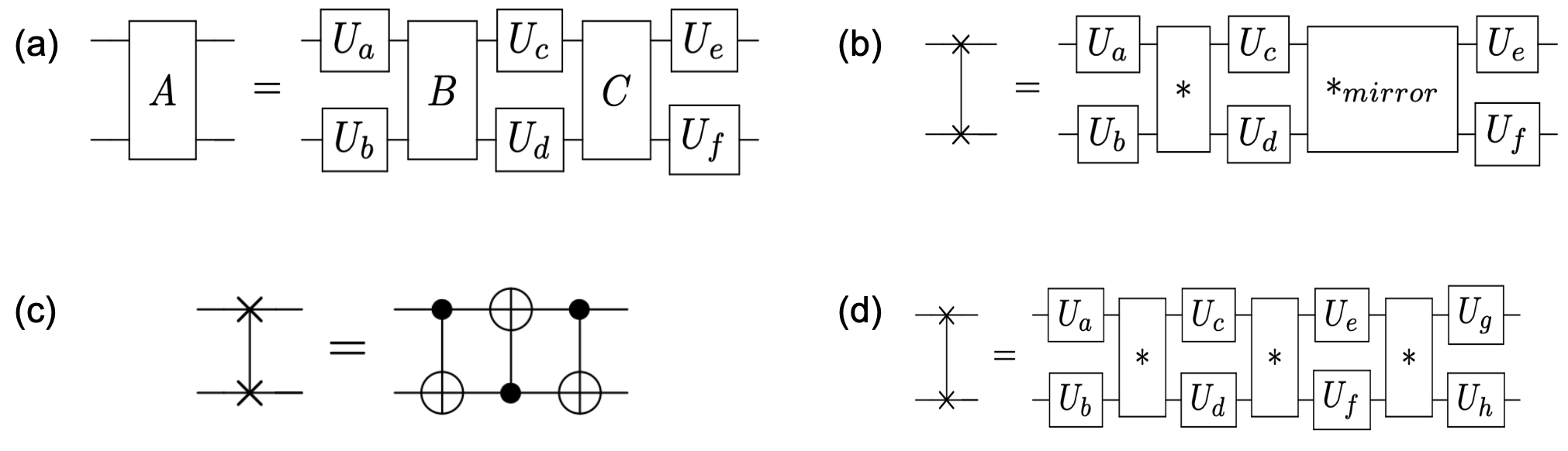}
    \caption{(a) Gate A, decomposed into 2 layers with 2Q gates B, C and 1Q gates $U_a$,$U_b$,$U_c$,$U_d$,$U_e$,$U_f$. (b) A general 2-layer decomposition of the SWAP gate. Here $\ast, \ast_{mirror}$ can be replaced by any pair of 2Q gates capable of synthesizing a SWAP in 2 layers. (c) The SWAP gate, decomposed into 3 CNOT gates. (d) A general 3-layer decomposition of the SWAP gate. Here the $\ast$ can be replaced by any 2Q gate capable of synthesizing a SWAP in 3 layers.}
    \label{fig:combined_circuits}
\end{figure*}

If a 2Q quantum gate is not directly supported on a device, it needs to be implemented by alternating layers of 1Q and 2Q gates from the set of basis gates that are directly supported. See Figure \ref{fig:combined_circuits} for examples. We say that a decomposition is $n$-layer if it contains $n$ layers of 2Q gates. Besides the errors that come from noises in the quantum hardware, a synthesized gate also suffers from the approximation error in gate decomposition. Thus the total fidelity of a gate should be the product of the hardware-limited fidelity and the decomposition fidelity. In this work, the decomposition errors are negligible compared to the hardware errors.

In our error model, decoherence is the dominant source of hardware error. So two factors determine whether a 2Q gate set is ideal for synthesizing a target gate: the duration of the basis gates, and the depth of the decomposition circuit. We need to take both into account when deciding on a strategy for selecting basis gates.

\subsection{An analytic method for determining 2Q circuit depth}
When deciding whether a potential basis gate is ideal for synthesizing a target gate, we consider the depth of the decomposition circuit as one of the factors. Given a 2Q target gate $A$, and a 2Q gate $B$ (or a gate set $S$), how to determine the minimum circuit depth required for a decomposition of A into $B$ (or $S$) and 1Q gates? One can take a practical, numerical approach to finding this decomposition. For a given number of layers, one can fix the 2Q gates and then numerically search for the 1Q gates that can minimize the discrepancy between the target unitary and the synthesized gate. One can start the numerical search from 1 layer, and increment the number of layers until the decomposition error gets below a threshold. But a more efficient and accurate way to determine the circuit depth is to apply the analytic method developed by Peterson et al. \cite{peterson}.

Without going into the technical details, here we summarize a key result from \cite{peterson} that we adapt and apply in Section \ref{swap_synthesis} and \ref{other_synthesis}.

\begin{thm}\label{peterson_thrm}
There exists a 2-layer decomposition of 2Q gate A into B, C, and 1Q gates as in Figure \ref{fig:combined_circuits}(a), if and only if any of the 1 to 8 sets of 72 inequalities that depend on the non-local parts of A, B, C is all satisfied.
\end{thm}

For details of the theorem, the readers can look at Theorem 23 of \cite{peterson} or the implementation of the function in our code \footnote{Our code can be found at \url{https://github.com/SophLin/nonstandard_2qbasis_gates}}. Note that Reference \cite{peterson} characterizes the space of 2Q gates with LogSpec instead of the Cartan coordinates. Both are valid ways to represent the non-local part of a 2Q gate, but care must be taken when converting between the two. A gate $U$ usually maps to 1 point in the Weyl chamber, but it usually maps to 2 points in the LogSpec space: $LogSpec(U) = (a,b,c,d)$ and $\rho(LogSpec(U)) = (c+\frac{1}{2},d+\frac{1}{2},a-\frac{1}{2},b-\frac{1}{2})$. If $LogSpec(U) = \rho(LogSpec(U))$ for all A, B, and C, we only need to check one set of inequalities. If $LogSpec(U) \neq LogSpec(U)$ for 1, 2, or all 3 of A, B, and C, we need to plug in different versions of the LogSpec and check 2, 4, or 8 versions of the 72 inequalities, respectively.

\begin{figure*}[!ht]
    \centering
    \includegraphics[width=\textwidth]{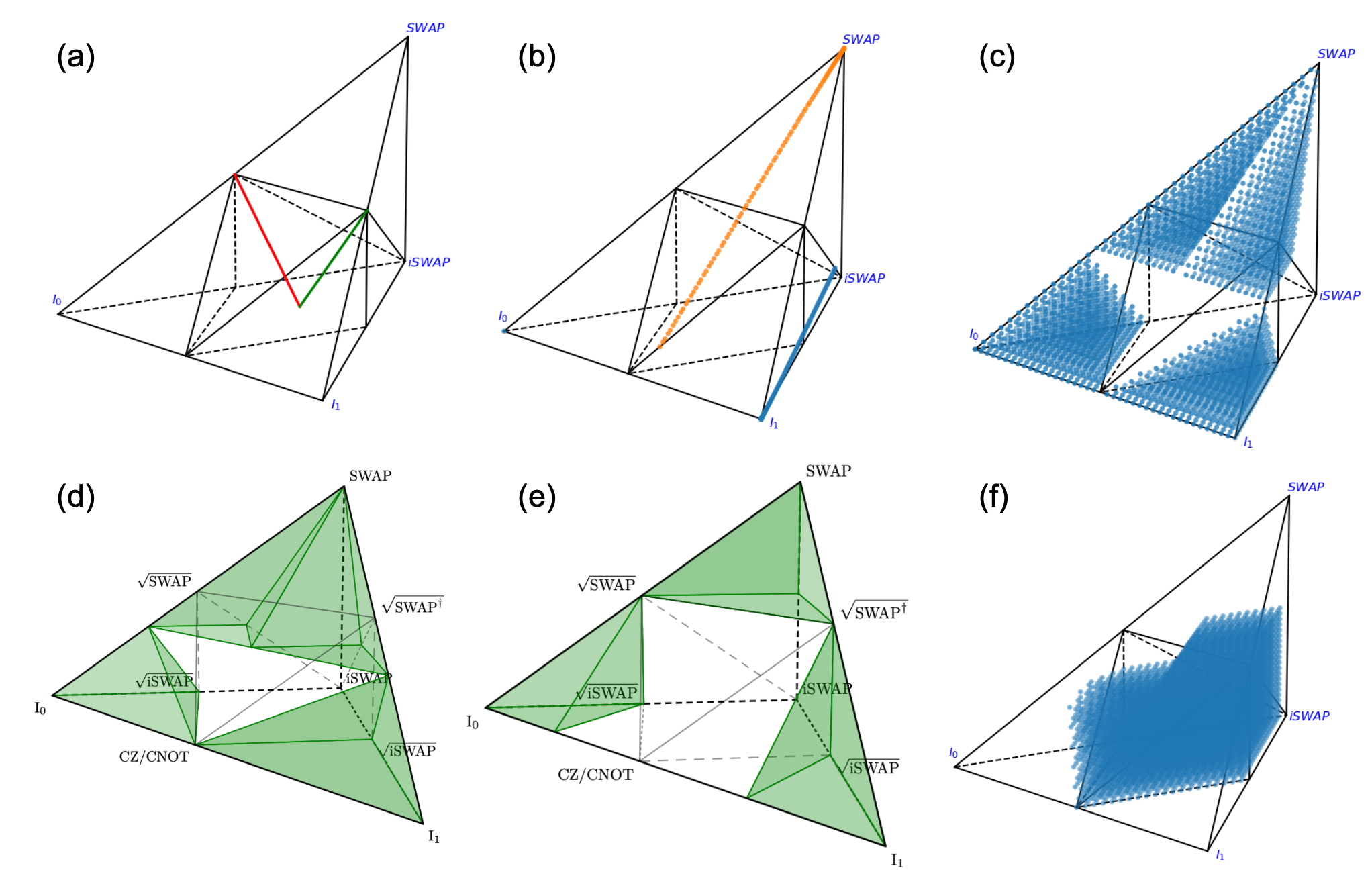}
    \caption{(a) Gates that are able to synthesize SWAP in 2 layers form 2 line segments in the Weyl chamber. The red one is from the B gate to $\sqrt{SWAP}$, and the green one is from the B gate to $\sqrt{SWAP}^\dagger$. (b) Pairs of gates that are able to synthesize a SWAP in 2 layers. In blue is an example trajectory that deviates from the standard XY interaction, in orange are the points that would complement the blue ones in synthesizing a SWAP in 2 layers. (c) Gates that are NOT able to synthesize a SWAP in 3 layers. (d) Gates that are NOT able to synthesize a SWAP in 3 layers. The 4 tetrahedra are defined by vertices $\{I_0, CZ, (\frac{1}{4}, \frac{1}{4},0), (\frac{1}{6},\frac{1}{6},\frac{1}{6})\}$, $\{CZ, I_1, (\frac{3}{4}, \frac{1}{4}, 0), (\frac{5}{6}, \frac{1}{6}, \frac{1}{6})\}$, $\{SWAP, (\frac{1}{2}, \frac{1}{6}, \frac{1}{6}), (\frac{1}{6}, \frac{1}{6}, \frac{1}{6}), (\frac{1}{3}, \frac{1}{3}, \frac{1}{6})\}$, and $\{SWAP, (\frac{1}{2}, \frac{1}{6}, \frac{1}{6}), (\frac{5}{6}, \frac{1}{6}, \frac{1}{6}), (\frac{2}{3}, \frac{1}{3}, \frac{1}{6})\}$. (e) Gates that are NOT able to synthesize CNOT in 2 layers. The 3 tetrahedra in the plot are defined by vertices $\{I_0, (\frac{1}{4},0,0), (\frac{1}{4},\frac{1}{4},\frac{1}{4}), \sqrt{SWAP}\}$, $\{I_1, (\frac{3}{4}, 0,0), (\frac{3}{4},\frac{1}{4}, 0), \sqrt{SWAP}^\dagger\}$, and $\{SWAP,\sqrt{SWAP}, \sqrt{SWAP}^\dagger, (\frac{1}{2}, \frac{1}{2}, \frac{1}{4}) \}$. (f) Gates that are able to decompose SWAP in 3 layers and CNOT in 2 layers.}
    \label{fig:combined_chambers}
\end{figure*}

\subsection{Synthesis of the SWAP gate}\label{swap_synthesis}
On bounded connectivity architectures, SWAPs make up a significant portion of all two-qubit gates. A SWAP gate exchanges the quantum states of two neighboring qubits. A 2Q gate in a quantum program can be directly scheduled if it acts on two physical qubits that are connected to each other, but this is not the case in general. Superconducting devices are usually designed to have sparse connectivity, because otherwise crosstalk errors would be difficult to suppress. As a result, quantum programs usually contain a large proportion of SWAP gates after they are compiled to run on a superconducting device. 

When we select the 2Q basis gate set for each pair of qubits, a top priority is to optimize the fidelity of the SWAP gate that is built from the gate set. 
We discuss three approaches towards synthesizing a SWAP gate: decompose it into 1, 2, or 3 layers of hardware 2Q gates.

\textbf{SWAP in 1 layer:} This requires a basis gate that is locally equivalent to SWAP. In other words, the trajectory of the available native gates needs to pass through the top vertex of the Weyl chamber.
   
\textbf{SWAP in 2 layers:} We consider 2 cases: 2-layer decomposition of SWAP using a single 2Q basis gate, and using two different 2Q basis gates.

In the first case, the set of 2Q gates that are capable of synthesizing SWAP in 2 layers are represented by 2 line segments in the Weyl chamber as shown in Figure \ref{fig:combined_circuits}(b). One is from the B gate to $\sqrt{SWAP}$ and the other is from $B$ to $\sqrt{SWAP}^\dagger$. We denote them by $L_0$ and $L_1$, respectively.

In the second case, for each point $\ast$ in the Weyl chamber, (as derived in Appendix B) there is exactly one point $\ast_{mirror}$ such that they together enable a 2-layer decomposition of SWAP (see Figure \ref{fig:combined_circuits}(b)). The line segment from $\ast$ to $\ast_{mirror}$ always has one of $L_0, L_1$ as its perpendicular bisector. Thus, given $\ast$, we can locate $\ast_{mirror}$ by rotating $\ast$ by $\pi$ around the closer one of $L_0, L_1$. One example pair of such points is CNOT and iSWAP. For a trajectory that deviates from the standard XY trajectory (goes from Identity to a point near iSWAP), its ``mirror'' is a trajectory from SWAP to a point near CNOT (Figure \ref{fig:combined_chambers}(b)). Since there's no overlap between the example trajectory and the ``mirror'', we conclude that the trajectory does not contain any pair of points that is able to synthesize SWAP together in 2 layers.

\textbf{SWAP in 3 layers:} It is a well-known result that 3 invocations of CNOT are required to implement a SWAP \cite{shende2003cnot3}. We show the circuit in Figure \ref{fig:combined_circuits}(c).
In fact, CNOT and iSWAP share the property that they can synthesize any arbitrary 2Q gate in 3 layers but only a $0$-volume set of gates (in the Weyl chamber) in 2 layers \cite{peterson}.

For our purpose, we need to know what other gates are capable of decomposing SWAP in 3 layers. We only consider 3-layer decomposition of SWAP using a single 2Q basis gate as in Figure \ref{fig:combined_circuits}(d). Let $S_{SWAP,3}$ denote the set of gates that satisfy our requirement. To determine whether a 2Q basis gate $G$ is in $S_{SWAP,3}$, we first locate the corresponding $G_{mirror}$ such that $G$ and $G_{mirror}$ together can provide a 2-layer decomposition of SWAP. Then we apply Theorem \ref{peterson_thrm} with $G_{mirror}$ as target and $G$ as basis gate to check if there exists a 2-layer decomposition of $G_{mirror}$ into $G$.

We apply the method above to a sample of points in the Weyl chamber, and obtain the distribution of gates that are able to synthesize SWAP in 3 layers. Since the complement of the set has a simpler shape, here we show a plot of $\overline{S_{SWAP,3}}$, the points that are not able to synthesize SWAP in 3 layers, in Figure \ref{fig:combined_chambers}(c). A visual inspection tells us $\overline{S_{SWAP,3}}$ consists of 4 tetrahedra in the Weyl chamber. After locating the vertices of the tetrahedra, we obtain Figure \ref{fig:combined_chambers}(d). We also learn that the volume of $S_{SWAP,3}$ is $68.5\%$ the volume of the Weyl chamber. 

A 2Q gate trajectory starts from either $I_0$ (or $I_1$) and goes out of the bottom left (or the bottom right) tetrahedron in Figure \ref{fig:combined_chambers}(d). If the trajectory does not go directly to SWAP, it will enter $S_{SWAP,3}$ after leaving the bottom tetrahedron that it starts from. Thus, the fastest gate on the trajectory that synthesizes SWAP in 3 layers can be found by locating the intersection of the trajectory with the face $\{CZ, (\frac{1}{4}, \frac{1}{4},0), (\frac{1}{6},\frac{1}{6},\frac{1}{6})\}$ or $\{CZ, (\frac{3}{4}, \frac{1}{4}, 0), (\frac{5}{6}, \frac{1}{6}, \frac{1}{6})\}$.

\textbf{Summary:} Given a 2Q gate trajectory that deviates from XY or XX, the most suitable 2Q gate for SWAP synthesis is the fastest one on the trajectory that is capable of synthesizing SWAP in 3 layers. Although some gates in the Weyl chamber are able to synthesize SWAP in 1 or 2 layers, it is unlikely that the early part of the trajectory overlaps any of them.


\subsection{Synthesis of other gates}\label{other_synthesis}
The techniques that we use to study the synthesis of SWAP also applies to other 2Q gates. For example, by applying Theorem \ref{peterson_thrm} to a sample of points in the Weyl chamber, with CNOT as target, we learn that the gates that are able to synthesize CNOT in 2 layers (denoted $S_{CNOT,2}$ here) takes up $75\%$ of the volume in the Weyl chamber. The complement $\overline{S_{CNOT,2}}$ consists of 3 tetrahedra, as shown in Figure \ref{fig:combined_chambers}(e). Therefore, on a 2Q gate trajectory, we can locate the fastest gate that synthesizes CNOT in 2 layers by taking the intersection of the trajectory with the face $\{(\frac{1}{4},0,0), (\frac{1}{4},\frac{1}{4},\frac{1}{4}), \sqrt{SWAP}\}$ or $\{(\frac{3}{4}, 0,0), (\frac{3}{4},\frac{1}{4}, 0), \sqrt{SWAP}^\dagger\}$. We can also locate the fastest gate from the trajectory that can both synthesize CNOT in 2 layers and synthesize SWAP in 3 layers, by taking the first intersection of the trajectory with $S_{CNOT,2} \cap S_{SWAP,3}$ (See Figure \ref{fig:combined_chambers}(f)).

\subsection{A strategy for locating good 2Q basis gates}\label{calibration_strategy}
Our framework allows one to prioritize different combinations of target 2Q gates. In Section \ref{case_study}, we test the following two criteria for selecting 2Q basis gates from native 2Q trajectories.
\begin{enumerate}
    \item Select the fastest gate on the trajectory that can synthesize SWAP in 3 layers.
    \item Select the fastest gate on the trajectory that can both synthesize SWAP in 3 layers and synthesize CNOT in 2 layers.
\end{enumerate}

As explained in Section \ref{swap_synthesis}, the gate that meets Criterion 1 can be found at the intersection of the 2Q trajectory and one of the 2 faces $\{CZ, (\frac{1}{4}, \frac{1}{4},0), (\frac{1}{6},\frac{1}{6},\frac{1}{6})\}$ and $\{CZ, (\frac{3}{4}, \frac{1}{4}, 0), (\frac{5}{6}, \frac{1}{6}, \frac{1}{6})\}$. And as explained in Section \ref{other_synthesis}, the gate that meets Criterion 2 can be found similarly. With this insight, we can locate a desired 2Q basis gate in an experimental setting using the methods in Section \ref{sec_calibration}.

Our framework can be easily adapted to other criteria for selecting basis gates. For instance, we can select the fastest gate that can decompose another set of target gates within a certain number of layers. We can also incorporate other metrics like the entangling power into a criterion, e.g. we can locate the fastest gate on the trajectory that is both a PE and can synthesize SWAP in 3 layers. 

\section{Calibration of nonstandard 2Q gates}\label{sec_calibration}

We propose two stages for calibrating a 2Q basis gate on an unknown trajectory of 2Q gates: first, a more costly ``initial tuneup'' stage that does not assume any knowledge of the trajectory and then a less costly ``retuning'' stage that utilizes information from the last initial tuneup and the retunings after it. In a well-controlled industry setup we would imagine the initial tuneup being done once a month and retuning being done daily. In a less well-controlled environment (e.g. one prone to low frequency drift), the initial tuneup could be done more frequently, as needed.

Our proposed calibration approach uses two techniques for experimentally characterizing the unitary of a potentially non-standard 2Q gate: quantum process tomography (QPT) \cite{qpt1,qpt2} and gate set tomography (GST) \cite{greenbaum2015_GST,nielsen2021_GST, Xue2022, Madzik2022}. QPT is a simple way to estimate a unitary but it cannot separate state preparation and measurement (SPAM) errors from gate errors \cite{qpt3}. GST is a highly general and accurate tomography technique that characterizes all the operations in a gate set (including SPAM) simultaneously and self-consistently.  
GST is simple to run, taking minutes to acquire on a superconducting device. GST acquisition is followed by classical processing of the data that can be computed on a cluster in about two hours. Note that during the classical processing, the quantum device can still be used with gates from the previous calibration cycle.

The speedup of GST's classical processing is an active field of research and may be obtained by allowing physics to inform the dominant errors that are expected \cite{pygsti}. The most relevant returns for fine tuning the unitary are the error generators \cite{Blume_Kohout_2022} for the gate set. The error generators are a basis for writing the transformation between the measured unitary and the unitary that GST expects. It measures coherent differences and estimates stochastic noise levels. GST is thus a valuable tool for directly characterizing 2Q gates.

Here we list the steps in the initial tuneup stage.

\begin{enumerate}

    \item Do preliminary coarse tuning experiments such as amplitude and frequency calibration of the entangling pulse drive to estimate the entangling pulse duration of interest. For example, a resonant $iSWAP$-like interaction may have an amplitude and a frequency to tune for optimal population swapping. (5 minutes per pulse)
    \item Perform QPT for each 2Q gate in the Cartan trajectory leading up to the approximate 2Q gate of interest. The qubit controller resolution (typically $\sim$1 ns) will determine the spacing between the trajectory points. Based on the findings in Step 1 the trajectory can be cropped around the entangling pulse duration of interest. The unitaries found will be the full list of candidate gates. (30-60 minutes per trajectory)
    \item From the candidate gates in the previous steps, use Section \ref{sec_theory} to identify which of them might be the fastest ones that also are good 2Q basis gates. In this step the list of candidate basis gates is narrowed down. We are not able to narrow down to one basis gate due to the imprecision of QPT.
    \item 
    Perform GST to obtain full information about each candidate 2Q gate, including an accurate gate unitary and a breakdown of error sources.
    Then the set of 2Q basis gates can be chosen. ($\sim$10 minutes for each 2Q gate, followed by classical processing)
\end{enumerate}

The second calibration stage is the quick ``retuning'' of the 2Q basis gates that relies on the results of the initial tuneup. Once the precise unitary for each 2Q basis gate is found, the gates can be simply retuned using the coarse tuning procedures in Step 1 of the initial tuneup. The information gained in the initial tuneup would allow experimentalists to prescribe a different retuning protocol to each 2Q basis gate according to what it needs. In practice, retuning would most likely be a simple combination of amplitude calibration and frequency calibration of the elements involved in each 2Q basis gate, and it would take approximately 1-5 minutes per 2Q basis gate.

\begin{figure}[!ht]
    \centering
    \includegraphics[width=0.99\columnwidth]{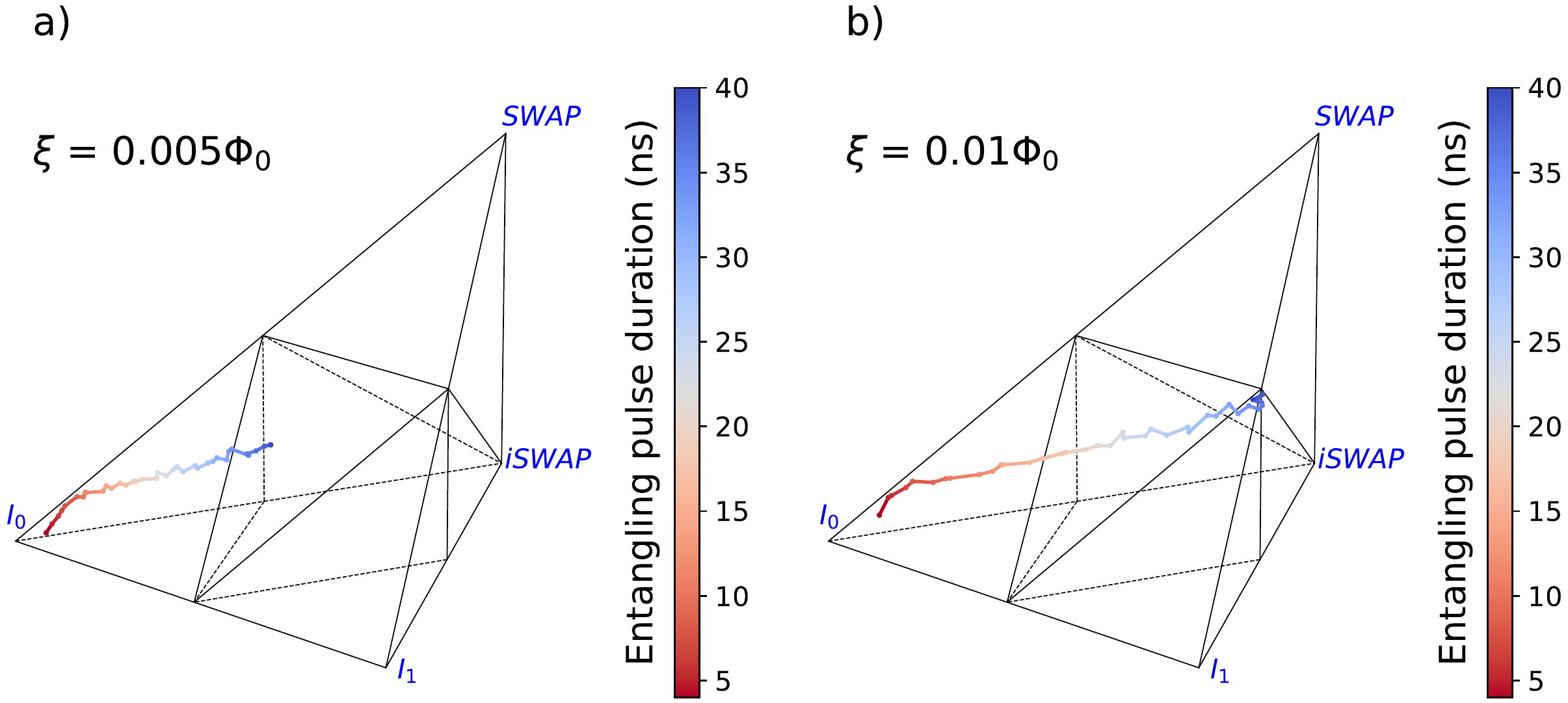}
    \caption{Stability over drive amplitude of the experimentally measured Cartan coordinate trajectories. In the same experimental implementation from Figure \ref{fig:Expt_Fig1}, as the entangling pulse drive amplitude $\xi$ increased from 0.005$\Phi_0$ to 0.01$\Phi_0$, the Cartan coordinate trajectories were found to double in speed but still be qualitatively similar. The data was collected over a two day period. As in Figure \ref{fig:Expt_Fig1}, due to an experimental hardware constraint the shortest possible entangling pulse duration was 4 ns, so the measured Cartan trajectories begin there.}
    \label{fig:Expt_Fig2}
\end{figure}

The extent to which previously gathered information can help reduce the cost of retuning depends on the stability of the gate trajectories over time. Figure \ref{fig:Expt_Fig2} shows the nonstandard Cartan trajectories measured on two days, over two entangling pulse drive amplitudes. Over the five day period that Cartan trajectories were measured for this device, the trajectories were all found to look qualitatively similar, as in Figure \ref{fig:Expt_Fig2}. While limited, this experimental data suggests that the measured Cartan trajectories obtained in the initial tuneup stage could potentially be used for several days afterward to provide an initial guess for the duration of the good 2Q basis gates. 
Our calibration protocol does not include the use of randomized benchmarking (RB) \cite{rb1,rb2,rb3}. RB is best suited for architectures with specific target gates that are members of the Clifford group. Furthermore, interleaved RB \cite{irb} will estimate the gate infidelity but will provide no information about an error budget. In our setting we do not have a fixed 2Q gate as the goal of implementation and understanding the gate unitaries themselves is a primary goal. We have thus decided GST and QPT are more suitable for precise gate characterization. 
The scalability of our proposed calibration method is not different from traditional approaches. Calibration techniques like QPT, RB, and GST can be applied to multiple 2Q gates on the same device in parallel, as long as the gates do not act on the same qubits. One can use an edge-coloring of the device connectivity graph to determine which gates to calibrate simultaneously. An edge-coloring of the grid graph takes 4 colors, one for a sparser connectivity (e.g. heavy hexagonal) takes fewer colors. Thus, for a superconducting device with typical connectivity, the calibration overhead on the quantum device does not scale with the size of the device.

\section{Compiling with non-standard 2Q basis gates}\label{sec_compile}

Most quantum programs and benchmarks are already specified at the 2 or 3 qubit gate level. Therefore, like previous works \cite{sqiswap}\cite{prakash}\cite{peterson2021xx} that discuss choice of 2Q basis gate and how to use less conventional 2Q basis gates for compilation, we use a transpiler pass to convert other 2Q gates in a circuit into our own 2Q basis gates, instead of building an entirely new compiler.

Some of the prior works decompose 2Q gates from application circuits into 1Q gates and native 2Q gates using a numerical approach \cite{prakash}, while others take an analytical approach \cite{sqiswap} \cite{peterson2021xx}. Note that such a decomposition requires finding the 1Q local unitaries, not just determining the required circuit depth. The analytical and numerical approaches each have their advantages. The numerical approach is more flexible. It can be applied to any 2Q basis and target gates. The analytic methods have limits on what gates they can be applied to, but are faster and some of them guarantee optimal results. There is currently no analytic formula that convert between arbitrary sets of 2Q gates. Huang et al. \cite{sqiswap} and Peterson et al. \cite{peterson2021xx} develop analytic algorithms that decomposes an arbitrary 2Q gates into $\sqrt{iSWAP}$ and discrete sets of XX-type gates, respectively. The {\small \texttt{decompose\_two\_qubit\_interaction\_into\_four\_ \\fsim\_gates}} function in Cirq \cite{Cirq} implements an analytic formula that decomposes an arbitrary 2Q gate into 4 layers of a given fSim gate, via the B gate.

In this project, we need to synthesize other 2Q gates from 2Q basis gates that are even less conventional than the ones considered in previous work. Therefore, we take a mostly numerical approach to gate synthesis and write our numerical search code based on NuOp from \cite{prakash}. The difference is, we use knowledge about decomposition circuit depth computed analytically to inform and speedup the numerical search for 1Q local unitaries. NuOp first attempts to search for a 1-layer decomposition, and moves on to 1 more layer upon failure to find solution, until it meets the target decomposition error rate. Using the analytic techniques for determining circuit depth developed by \cite{peterson} and extended by our work for SWAP, we are able to skip to the step in NuOp in which a perfect decomposition is guaranteed by theory. This significantly speeds up the numerical search and also guarantee that the solution has optimal depth.

Synthesizing all 2Q gates in the application programs directly into the basis gates might incur a compilation overhead. We avoid it by computing in advance and storing the decompositions of a few common 2Q gates into our basis gates. This only needs to be done once per calibration cycle (usually 1 day) and costs little time. In this work (see Section \ref{case_study}) we only directly decompose SWAP and CNOT into our basis gates. But instead of taking this minimalist approach, one can alternatively prepare decompositions for a larger set of potential target gates into the basis gates. The cost would still quite small. We imagine that one can identify a set of potentially useful target gates using an approach similar to \cite{prakash}, except that \cite{prakash} looks for a set of gates to calibrate instead of decompose. In addition, in the scenario where programs wait in long queues before execution, one might be able to afford directly decomposing all 2Q gates in the circuits into the basis gates.

\section{Case study: entangling fixed \\ frequency far-detuned \\ transmons with a tunable \\ coupler}\label{case_study}

\subsection{Introduction to the case study entangling gate architecture}

 Many efforts are being made in industry and academia to design a 2Q entangling gate architecture that can be used for scaling up to a general quantum computer \cite{ibmCR,foxen,sung2020realization}. The all-microwave cross-resonance gate was recently used by IBM to do a high fidelity CNOT gate in 90 ns \cite{ibmCR}, but to suppress the always-on ZZ crosstalk mentioned in Section \ref{deviation}, precise crosstalk cancellation pulses applied to both qubits during run time were required, adding complexity to the architecture. Google Quantum AI and MIT have each developed entangling gate architectures for high fidelity CZ and iSWAP gates, with Google's architecture supporting a continuous set of these standard gates \cite{foxen,sung2020realization}. Google's architecture requires all qubits and couplers to be flux-tunable, which adds complexity and additional sources of leakage and noise to their architecture. Similarly, in order to suppress the always-on ZZ crosstalk, MIT's architecture requires one qubit per pair to be tunable as well as the coupler.

The unit cell of our case study entangling gate architecture is a pair of qubits and a coupler. This unit cell, first proposed in \cite{hamiltonian_source}, was designed to perform a diverse set of 2Q gates, including iSWAP and CZ; the full list of 2Q gates can be found in Table 1 of \cite{hamiltonian_source}. The two qubits are fixed frequency transmon qubits; the benefits of fixed frequency transmons are that they are easy to fabricate and can be reliably engineered to have high coherence $> 100$ us \cite{place}. The two qubits are also far detuned from each other so there is reduced single qubit control crosstalk. The coupler is a generalized flux qubit which has been designed to have several good properties for qubit control. Notably, because the coupler's positive anharmonicity has been designed to balance out the negative anharmonicity of the two qubits, the eigenspectrum of this architecture's unit cell can support a zero-ZZ crosstalk bias point. This architecture is relatively simple to implement because fixed frequency transmons have high coherence, there is only one flux-tunable element in the unit cell (the coupler), and it is easy to bias the unit cell to zero-ZZ crosstalk.

A model Hamiltonian of the two qubits coupled with a tunable coupler is shown in Appendix A. Here we highlight the time-dependent term, $\hat{H}_c(t)$ that describes the coupler dynamics:
\begin{equation}
     \hat{H}_c(t) = \omega_c(t) \hat{c}^\dagger \hat{c} + \frac{\alpha_c}{2} \hat{c}^{\dagger 2} \hat{c}^2
     \label{eqn:hamiltonian}
\end{equation}
where $\alpha_c$ is the coupler anharmonicity, $\hat{c}$ is the annihilation operator and the coupler frequency $\omega_c(t)$, corresponding to the transition to its first excited state, can be varied in time via the flux through its superconducting loop. Low-crosstalk 2Q gates are realized by AC modulating this coupler frequency after DC biasing it to the zero-ZZ crosstalk bias point.

\begin{figure}
    \centering
    \includegraphics[scale=0.4]{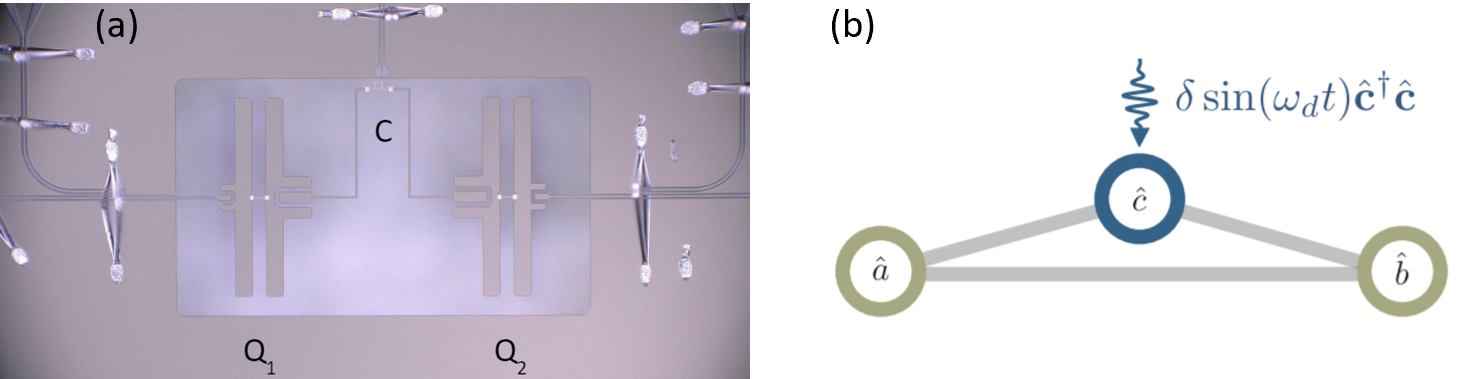}
    \caption{(a) Optical image of the device presented in \cite{experimental_work} shows two fixed frequency transmons coupled via a tunable coupler. (b) Schematic for modelling the device adapted from \cite{hamiltonian_source}.} 
    \label{fig:device}
\end{figure}

In \cite{experimental_work} an early prototype device (shown in Fig. \ref{fig:device}) for this case study architecture demonstrated a fast perfect entangler biased to zero-ZZ crosstalk. This device produced the nonstandard 2Q gate trajectory shown in Figure \ref{fig:Expt_Fig1}, which included a 13 ns perfect entangler. Figure \ref{fig:Expt_Fig2} shows how the measured trajectories were similar over a range of entangling pulse drive amplitudes that did not exceed $\xi =$ 0.01$\Phi_0$, the point at which strong drive effects would be expected to become non-negligible \cite{hamiltonian_source}. So in this early prototype device, the measured trajectories in Figures \ref{fig:Expt_Fig1} and \ref{fig:Expt_Fig2} were not nonstandard because of strong drive effects, but because of some other systematic in the experiment.

\subsection{Our simulation approach}

The case study entangling gate architecture natively supports strong parametrically activated interactions between the two qubits. Since the full Hamiltonian for this architecture is computationally intensive to model \cite{hamiltonian_source}, for our simulation we use the simplified effective Hamiltonian from \cite{hamiltonian_source} that models the device using fewer parameters while still capturing all of the essential physics of the device (see Appendix A). Our general protocol for simulating Cartan trajectories is as follows:

\begin{enumerate}
    \item We input the simulated device parameters into our Hamiltonian. These parameters include the qubit frequencies $\omega_a$ and $\omega_b$, and the qubit coherence times. This generates the eigenspectrum of the simulated device.
    \item We bias the coupler frequency ($\omega_c^0$) between the two qubit frequencies ($\omega_a,\omega_b$) such that the static ZZ term (i.e. for $\delta(t)=0$) between the two qubits is tuned to zero. 
    \item We specify the drive amplitude $\xi$ of our entangling pulse. In this case study we implement a iSWAP-like entangler, so the entangling pulse is driven at the frequency $\omega_d$ that generates maximal population swapping between the two qubits. For $\xi \leq $ 0.01$\Phi_0$, the entangling pulse frequency $\omega_d$ is essentially identical to the difference frequency of the two qubits $|\omega_a-\omega_b|$. However, increasing $\xi > 0.01\Phi_0$ activates the two-photon process in Equation \ref{eqn:hamiltonian}, causing population to enter the second excited state of the coupler and modify the entangling interaction. This in turn causes $\omega_d$ to deviate from $|\omega_a-\omega_b|$. The entangling pulse is modulated by a rectangular envelope, as was done in experiment to obtain the measured trajectories; due to qubit controllers typically having a time resolution of 1 ns, short entangling gates $\sim$10 ns have to be implemented using a pulse with a fast rise time. Experimentalists typically choose between a flat top Gaussian pulse with a short rise time, or a rectangular pulse for simplicity. 
    \item We evolve the time-dependent Hamiltonian and project the evolution propagator on the computational subspace to obtain the effective unitary operation with respect to the entangling pulse drive duration. This time ordered sequence of unitary operations can be represented as a trajectory in the Weyl space using Cartan coordinates. By examining the trace of the effective unitary propagator we can obtain the leakage outside the computational space. We confirm that the leakage rates are much below the expected gate errors due to decoherence. 
\end{enumerate}

In this case study we simulate standard and nonstandard 2Q trajectories. The simplest and most consistent way to do this is to use the same simulated devices but to vary the drive power $\xi$. For $\xi \leq $ 0.01$\Phi_0$ we expect the above protocol to result in a standard iSWAP interaction between the two qubits. But for $\xi > 0.01\Phi_0$, we expect strong drive effects to begin to emerge and cause the Cartan trajectory to deviate away from a standard iSWAP. We note that the simulated trajectories differ in several ways from the measured trajectories in Figures \ref{fig:Expt_Fig1} and \ref{fig:Expt_Fig2}. Firstly, the measured trajectories are nonstandard even for $\xi \leq $ 0.01$\Phi_0$ due to an additional systematic effect in the experiment. Secondly, the simulated trajectories are consistently slower than the measured trajectories; e.g. at $\xi$ = 0.01$\Phi_0$, the simulated trajectories are slower by a factor of 3.5 than the measured trajectory, which included a 13 ns $\sqrt{iSWAP}$-like entangling gate. These discrepancies can both be explained by the simulation model Hamiltonian being significantly simpler than the true device Hamiltonian. Aside from these discrepancies, our simulations are realistic; our trajectories are generated using parameters and techniques that closely resemble those used in experiment and our method for generating standard and nonstandard trajectories using a single simulated device is physically intuitive. 

Simulating Cartan trajectories over a range of entangling pulse amplitudes $\xi$ we observe the correct intuitive behavior. The simulated trajectories deviate more and more from the standard iSWAP as the entangling pulse amplitude increases beyond $\xi$ = 0.01$\Phi_0$. Secondly, the speed of the simulated trajectories scales linearly with $\xi$. This agrees with the experimental data shown in Figure \ref{fig:Expt_Fig2} where the measured trajectory doubled in speed when $\xi$ increased by a factor of two.

\subsection{Methodology}
\begin{figure}
    \centering
    \input{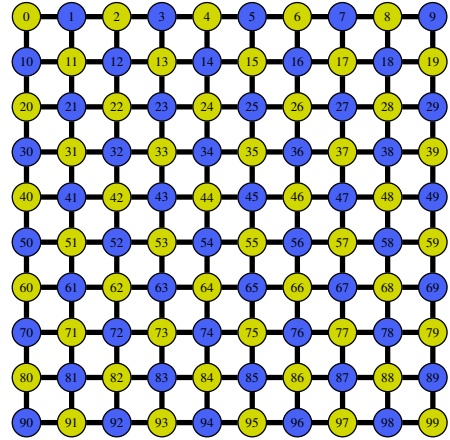}
    \caption{Device simulation. The high and low frequency qubits are shown in different colors. Each edge connects two qubits with different colors.}
    \label{fig:connectivity}
\end{figure}

We simulate a 10 by 10 device with grid connectivity (Fig. \ref{fig:connectivity}),where the qubit frequencies of each pair of neighbors are sampled from two normal distributions respectively with means that differ by 2 GHz.
We use a $5\%$ standard deviation for sampling the qubit frequencies. Improved fabrication techniques have reduced the smaller standard deviation to about $0.5\%$ \cite{qubit_freq_std}, but we use a larger standard deviation to show that our method is robust to variations in device fabrication.

Between each pair of neighboring qubits on the $10\times 10$ grid, we simulate two types of 2Q trajectory by varying the entangling pulse amplitude $\xi$: 1) A baseline trajectory generated with a low entangling pulse amplitude of $\xi$ = 0.005$\Phi_0$ and 2) a nonstandard trajectory due to strong drive effects resulting from a larger $\xi$ = 0.04$\Phi_0$.

Then on each nonstandard trajectory, we select 2Q basis gates using Criterion 1 and 2 (respectively) introduced in Section \ref{calibration_strategy}. We test these 3 sets of 2Q basis gates on common application circuits as benchmarks. We use the Qiskit\cite{Qiskit} transpiler with the ``SABRE''\cite{sabre} layout and routing methods to map the benchmarks circuits to the $10\times 10$ grid connectivity. With the nonstandard basis gates, we compile circuits using the methods from Section \ref{sec_compile}. With the $\sqrt{iSWAP}$ from the standard trajectories, we use the analytic approach in \cite{sqiswap}. Like the 2Q basis gates selected with Criterion 2, $\sqrt{iSWAP}$ decomposes SWAP in 3 layers and CNOT in 2 layers, but we can also use it to directly decompose other 2Q gates (like the CRZ gates in the QFT benchmarks) analytically. For the 1Q gates in the gate and circuit synthesis, we use a duration of $20$ ns, which is typical for fixed-frequency transmon qubit processors \cite{jurcevic2021demonstration}. 

Decoherence is the dominant hardware noise in our noise model, because crosstalk is suppressed by the high detuning in the qubits. For each qubit, we model the decoherence error as $1-e^{-t/T}$, where $T$ is the coherence time of the qubit. We set $T$ to a typical value of 80 $\mu$s for all qubits. We compute $t$ as $t_f - t_i$, where $t_i$ is the start of the first gate on the qubit and $t_f$ is the end of the last gate on the qubit. The total coherence-limited fidelity of a circuit is the product over the $e^{-t/T}$ term from each qubit. The decomposition errors in gate synthesis are negligible compared to the decoherence errors, and can be reduced to arbitrarily close to zero in theory. Thus we only show the coherence-limited fidelities in the results.

\subsection{Results}

Before discussing our results, as a disclaimer we note that while increasing the entangling pulse drive amplitude is one way to speed up 2Q gates, it is by no means an all-purpose solution that we generally advocate for. We chose to do this in our simulation case study only because it was a simple and intuitive way to compare standard and nonstandard simulated gates for the same case study entangling architecture. For this case study architecture, the drive amplitudes chosen were realistic in an experimental setting.
\begin{table}[!h]
\centering
\renewcommand{\arraystretch}{1.5}
\resizebox{0.38\textwidth}{!}{
\begin{tabular}{|l|r|r|r|}
\hline
  & Basis & SWAP & CNOT \\ [0.5ex] 
\hline
\multirow{2}{*}{Baseline} & 83.04 ns & 329.1 ns & 226.1 ns\\    
& 99.884\% & 99.541\% & 99.684\% \\ 
\hline
\multirow{2}{*}{Criterion 1} & 10.15 ns & 110.5 ns & 110.5 ns\\    
& 99.986\% & 99.845\% & 99.845\% \\
\hline
\multirow{2}{*}{Criterion 2} & 10.76 ns & 112.3 ns & 81.51 ns\\    
& 99.985\% & 99.843\% & 99.886\% \\
\hline
\end{tabular}}

\caption{Average duration (top) and coherence-limited gate fidelity (bottom) of the 2Q basis gates and the synthesized SWAP and CNOT gates, from baseline, Criterion 1, and Criterion 2.}
\label{table:gate_duration}
\end{table}

\begin{table}[!b]
\begin{tabular}{|l|r|r|r|}
\hline
Benchmark     & \multicolumn{1}{l|}{Baseline } & \multicolumn{1}{l|}{Criterion 1} & \multicolumn{1}{l|}{Criterion 2} \\ \hline
qft 10        & $58.2\%$                                & $65.6\%$                                    & $70.8\%$                                    \\ \hline
qft 20        & $1.33\%$   & $6.03\%$    &   $9.94\%$                                \\ \hline
bv 9          &   $88.7\%$                            &   $94.4\%$     &    $95.3\%$                                \\ \hline
bv 19         &   $79.3\%$                &  $89.9\%$           &  $91.0\%$\\ \hline
bv 29         &   $44.5\%$         &   $72.5\%$     &  $74.3\%$   \\ \hline
bv 39         &  $26.8\%$         &    $56.3\%$            & $59.7\%$                \\ \hline
bv 49         &  $27.7\%$          &  $58.4\%$               &  $62.4\%$     \\ \hline
bv 59         &   $12.5\%$                              &     $43.8\%$                               &  $47.4\%$                                   \\ \hline
bv 69         &  $9.15\%$           &   $39.4\%$                               & $43.2\%$                                   \\ \hline
bv 79         &  $0.428\%$                             &  $11.3\%$                                 &      $14.2\%$                              \\ \hline
bv 89         & $2.44\%$                              &  $23.1\%$                                   &   $26.3\%$                                 \\ \hline
bv 99         &    $0.06\%$                        &    $6.26\%$                            &       $7.97\%$                            \\ \hline
cuccaro 10    &     $21.5\%$                           &    $46.3\%$                                & $52.6\%$                                   \\ \hline
cuccaro 20    &  $0.800\%$                            &     $7.68\%$                               &    $11.8\%$                               \\ \hline
qaoa\_0.1 10  &    $97.2\%$  &  $98.5\%$   &  $98.8\%$          \\ \hline
qaoa\_0.1 20  &  $84.4\%$      &  $92.0\%$      &    $93.6\%$                            \\ \hline
qaoa\_0.1 30  &    $14.4\%$   &  $43.3\%$        &   $49.0\%$                      \\ \hline
qaoa\_0.1 40  &  $0.00585\%$                         &  $5.59\%$    & $8.56\%$                            \\ \hline
qaoa\_0.33 10 &   $66.1\%$   &   $81.0\%$          & $84.3\%$    \\ \hline
qaoa\_0.33 20 &   $15.0\%$    &  $42.2\%$      &  $48.2\%$      \\ \hline
\end{tabular}
\caption{The decoherence-limited fidelities of benchmark circuits, transpiled using the standard 2Q basis gates from baseline, and the nonstandard ones selected by Criterion 1 and 2. The QAOA benchmarks all have $p=1$ where $p$ is the number of times the protocol is repeated. The fractions 0.1 and 0.33 are the probablities that an edge is created between a pair of nodes.}
\label{table:circuit_fid_results}
\end{table}

The average durations and coherence limited fidelities (obtained using the Qiskit Ignis \texttt{coherence\_limit} function \cite{ignis}) of the synthesized SWAP and CNOT gates from the two approaches are summarized in Table \ref{table:gate_duration}. In Table \ref{table:circuit_fid_results}, we show the coherence-limited circuit fidelities of 5 sets of benchmark circuits, when transpiled to different sets of 2Q basis gates.  
We first observe that the faster nonstandard 2Q basis gates have $\sim$8x lower coherence-limited infidelities than the baseline standard 2Q gates. We also observe that the synthesized SWAP (CNOT) gates from Criterion 1 and 2 are 3.0x and 2.9x (2.0x and 2.8x) faster than the baseline, respectively. Due to the relation between gate fidelity and circuit fidelity, fidelity improvements scale exponentially in benchmark size.

Next, we observe that Criterion 2 performs better than Criterion 1. This is not surprising since it has significantly faster CNOT gates and only slightly slower SWAP gates compared to Criterion 1.

For the baseline case, the 1Q gate duration is 4x shorter than the standard 2Q basis gate, and therefore $\sim$24\% of the duration of the compiled SWAP/CNOT gate is spent performing 1Q gates. In contrast, for the nonstandard case, the 1Q gate duration is 2x longer than the nonstandard 2Q basis gate, and $\sim$72\% of the duration of the compiled SWAP/CNOT gate is spent performing 1Q gates. This puts us in the regime of today’s fastest large superconducting processors such as Google's Sycamore device, where the optimal processor configuration that minimizes the overall effects of gate error has the 1Q gates being roughly twice as long as the 2Q gates \cite{Arute2019}.

\section{Conclusion}\label{sec_conclusion}

The idea of a uniform set of basis gates naturally arose from early notions of universal gate sets, which experimentalists then implemented on various qubit platforms.  By looking at the theory of possible entanglers, we have found that there are many options for good 2Q basis gates, and that these gates behave differently on each pair of interacting qubits in a processor. This led us to a radically new idea, why be constrained to a single canonical gate (e.g. CX or CZ)?  Why not tune up the gate that will have the highest fidelity between every pair of qubits, allowing each to differ and instead adjust for these variations in software? If we do not treat all the coherent deviations in gate trajectories as errors, we will have more freedom in hardware design and achieve a higher gate fidelity.

In this paper, we examined the space of possible entanglers and developed a practical method for finding a high-fidelity entangler between every pair of qubits. In the case study, we find heterogeneous basis gates that are $\sim$8x faster than the baseline, and use them to synthesize faster SWAP and CNOT gates than synthesized by the baseline $\sqrt{iSWAP}$ gate from the standard XY-type trajectories.
We then evaluate these heterogeneous basis gates on a number of benchmark circuits and find fidelity improvements that scale exponentially in benchmark size.

Our approach successfully uses software to overcome the limitations of today's hardware. Such types of adaptive basis-gate design will be essential to pioneering innovative future quantum systems.

\section*{ACKNOWLEDGMENTS}
We would like to thank Charlie Guinn for discussions about calibration and tomography techniques.

This work is funded in part by EPiQC, an NSF Expedition
in Computing, under grants CCF-1730082/1730449; in part
by STAQ under grant NSF Phy-1818914; in part by NSF
Grant No. 2110860; by the US Department of Energy Office 
of Advanced Scientific Computing Research, Accelerated 
Research for Quantum Computing Program; and in part by 
NSF OMA-2016136 and the Q-NEXT DOE NQI Center. This work was completed in part with resources provided by the University of Chicago’s Research Computing Center. SS is supported by the Department of Defense (DoD) through the National Defense Science \& Engineering Graduate Fellowship (NDSEG) Program. FTC is Chief Scientist at Super.tech and an
advisor to Quantum Circuits, Inc.

\appendices
\section{Hamiltonian of 2 qubits coupled with a tunable coupler}
The system Hamiltonian of the two qubits coupled with a tunable coupler can be modelled as in \cite{hamiltonian_source}:

\begin{align}\label{Eq:Model}
 \hat{H}(t) = \hat{H}_a + \hat{H}_b + \hat{H}_c(t) + \hat{H}_g,
\end{align}

with

\begin{align}
  \begin{split}
  \hat{H}_a &= \omega_a \hat{a}^\dagger \hat{a} + \frac{\alpha_a}{2} \hat{a}^{\dagger 2} \hat{a}^2,  \\
  \hat{H}_b &= \omega_b \hat{b}^\dagger \hat{b} + \frac{\alpha_b}{2} \hat{b}^{\dagger 2} \hat{b}^2,  \\
  \hat{H}_c(t) &= \omega_c(t) \hat{c}^\dagger \hat{c} + \frac{\alpha_c}{2} \hat{c}^{\dagger 2} \hat{c}^2.  \\
  \hat{H}_g    &= -{g}_{ab} \hat{a}^\dagger \hat{b} - {g}_{bc} \hat{b}^\dagger \hat{c} - {g}_{ca} \hat{c}^\dagger \hat{a}\\ & -{g}_{ab}^* \hat{a} \hat{b}^\dagger - {g}_{bc}^* \hat{b} \hat{c}^\dagger - {g}_{ca}^* \hat{c} \hat{a}^\dagger  
  \end{split} 
\end{align}

where $\omega_{a(b)}$ corresponds to the qubit a(b) frequency, $g_{ij}$ represents capacitive coupling strength between elements $i$ and $j$. The entangling interaction is realized by modulating the coupler frequency as $\omega_c(t) = \omega_c^0 + \delta \sin(\omega_d t)$.

\section{SWAP synthesis in 2 layers}
See the circuit in Fig. \ref{fig:combined_circuits}(a). Let $A=SWAP$ we get the equation $$SWAP = (e\otimes f) C (c\otimes d) B (a\otimes b).$$
Move $e\otimes f$ and $a\otimes b$ to the other side and move $e\otimes f$ through SWAP,
\begin{align*}
    C (c\otimes d) B &=(e\otimes f)^\dagger SWAP (a\otimes b)^\dagger\\
    &= SWAP (f\otimes e )^\dagger (a \otimes b)^\dagger\\
    &= SWAP (fa\otimes eb )^\dagger.
\end{align*}
Move $(fa\otimes eb )^\dagger$ to the LHS, and C to the RHS,
$$(c\otimes d) B (fa\otimes eb) = C^\dagger SWAP.$$
This equation tells us that, $B$ and $C$ can synthesize SWAP as in Fig. \ref{fig:combined_circuits}(a) if and only if the Cartan coordinates of $B$ are equal to the Cartan coordinates of $C^\dagger SWAP$ up to canonicalization. Let $B \sim (x,y,z)$ and $C \sim (x',y',z')$, then we have $(x,y,z) \sim (-x',-y',-z') + (\frac{1}{2}, \frac{1}{2}, \frac{1}{2})$. From this we can tell that for every local equivalence class $[B]$ of 2Q gates, there is exactly one local equivalence class $[C]$ such that $[B]$ and $[C]$ together can synthesize SWAP in 2 layers. And since we know how to canonicalize Cartan coordinates into points within the Weyl chamber, given $[B]$ we will be able to find the corresponding $[C]$. Here we do not elaborate on how we identify the geometric relation between $[B]$ and $[C]$ inside the Weyl chamber, but the readers can check our claim by applying Theorem \ref{peterson_thrm}.


\bibliographystyle{IEEEtran}
\bibliography{main}

\end{document}